\documentclass[showpacs,twocolumn]{revtex4}
\usepackage{amsmath,amssymb,mathrsfs,latexsym,graphicx,psfrag}
\newcommand{\bs}[1]{\boldsymbol{#1}}

\newcommand{\ket}[1]{\left|#1\right\rangle}

\newcommand{\up}{\uparrow}
\newcommand{\dw}{\downarrow}

\def\ie{\emph{i.e.},\ }
\def\eg{\emph{e.g.}\ }

\allowdisplaybreaks[1]
\begin{document}
\title{Many-spinon states and the secret significance of Young tableaux}
\author{Martin Greiter and Dirk Schuricht}
\affiliation{Institut f\"ur Theorie der Kondensierten Materie,
  Universit\"at Karlsruhe,\\ Postfach 6980, 76128 Karlsruhe, Germany}
\pagestyle{plain}
\begin{abstract}
  We establish a one-to-one correspondence between the Young tableaux
  classifying the total spin representations of $N$ spins and the
  exact eigenstates of the the Haldane-Shastry model for a chain with
  $N$ sites classified by the total spins and the fractionally spaced
  single-particle momenta of the spinons.
\end{abstract}
\pacs{75.10.Pq, 02.30.Ik, 05.30.Pr, 02.20.-a\vspace{-2pt}}
%
%
%
\maketitle

Integrable models~\cite{KorepinBogoliubovIzergin93,andrei92proc} have
always played a special role in our understanding of interacting
many-particle systems in one dimension.  While it is clearly
impossible to realize an integrable model experimentally, it is clear
that many concepts and mechanisms discovered through the study of
integrable systems are directly or indirectly relevant to a
substantial body of experimental data.  An infinite set of conserved
quantities renders many integrable models amenable to exact solution
through one or several variations of the Bethe ansatz method, and
hence allows us to get a grasp on the non-trivial physical concepts
involved.  The most prominent among these concepts is probably the
fractional quantization of spin in antiferromagnetic spin $1/2$
chains.  Faddeev and Takhtajan~\cite{faddeev-81pla375} discovered in
1981 that the elementary excitations (now called spinons) of the spin
$1/2$ Heisenberg chain solved by Bethe~\cite{bethe31zp205} in 1931
carry spin $1/2$ while the Hilbert space for the spin chain is spanned
by spin flips, which carry spin 1.  The fractional quantization of
spin in spin chains is conceptually very similar to the fractional
quantization of charge in quantized Hall
liquids~\cite{laughlin83prl1395, stone92}.  In the case of the
Haldane-Shastry model (HSM)~\cite{haldane88prl635,
  shastry88prl639,haldane91prl1529,kawakami92prb1005,
  haldane-92prl2021}, which we will elaborate on below, the analogy
even extends to the explicit wave functions for the ground states and
the spinon or quasihole excitations of the spin chain and the Hall
liquid, respectively.

The discovery of the spinon through the Bethe ansatz (BA) solution
illustrates both the importance of integrable models and BA
techniques as well as the practical limitations, as it took 50 years
to read off as elementary a property as the spin of an excitation in a
known and established exact solution to the problem.  The BA is not a very
practical method to calculate observable quantities, both because the
number of integrable models is limited and because it is often
exceedingly difficult to extract physical quantities like correlation
or response functions from the exact solutions.  One of the reasons
underlying these practical limitations may be that the BA
solutions are given as distributions of pseudomomenta, in which spinon
excitations appear as defects.  Spinons hence play the role of defects
or solitons in the solutions of the BA equations, just as
they and the quasiparticles in quantized Hall liquids are often viewed
as solitons or collective excitations in states constructed of spin
flips or electrons, respectively.  In many regards, however, spinons
can and should be interpreted as particles, which requires solutions
labelled in terms of spinon quantum numbers.  Such solutions would, in
principle, allow for the development of perturbative methods in the
spinon fields directly.

With regard to this perspective, there is encouragement and there are
problems.  The encouraging news is that the HSM provides us with an
ideal starting point for any such perturbative expansion, since the
spinons in this model are free~\cite{ha-93prb12459,
  essler95prb13357,greiter-05prb224424} in the sense that they
interact only through their half-Fermi statistics.  The half-Fermi
statistics yields non-trivial state counting
rules~\cite{haldane91prl937, schoutens97prl2608, bouwknegt-99npb501}
and fractional momentum spacings~\cite{greiter06}.  It is still far
from simple to construct the spinon Hilbert space, as it cannot be
decomposed into a product space of single-particle states, known as
Fock space in the familiar cases of fermions or bosons.  
In this Letter, we propose what we believe to be the simplest construction.

The main problem with the development of perturbative methods in terms
of spinon fields is that the matrix elements of local spin operators
$\bs{S}_{\alpha}$ on sites $\alpha$ of the chain between states with
different numbers of free spinons, \ie eigenstates of the HSM, are
required.  At present, only very few of these matrix elements are
known, and the exact expressions for these elements for finite chains
appear rather complicated~\cite{yamamoto-00prl1308, 
  bernevig-01prb24425,peysson03jpa7233}.  
These expressions, however, greatly simplify in the thermodynamic
limit, and there is hope that a method to obtain them directly in this
limit can be developed.

Let us now turn to the Hilbert space representation for spinons in an
SU(2) or, in general, SU($n$) spin chain.  The non-trivial feature is
that the single-spinon quantum numbers depend on the occupations of
all the other single-spinon states.  These quantum numbers are the
spins and the single-particle momenta of the spinons.  It would
probably be exceedingly difficult to obtain them if the
Haldane-Shastry model would not provide us with a framework.  Since
the spinons in this model are free, the problem of finding the
equivalent of a Fock space representation for spinons reduces to
finding a convenient representation of the eigenstates of this model
in terms of their spinon content.  In other words, we set out to find
a general set of rules to determine the allowed sequences of
single-spinon momenta $p_1,\ldots,p_L$ as well as the allowed
representations for the total spin of the states such that the
eigenstates of the HSM have momenta and energies
\begin{equation}
  p=p_0+\sum_{i=1}^L p_i,\quad E=E_0+\sum_{i=1}^L \epsilon(p_i),
  \label{eq:Lspinonenergy}
\end{equation} 
where $p_0$ and $E_0$ denote the ground state momentum and energy,
respectively, and $\epsilon(p)$ a single-spinon dispersion irrelevant
to the general purpose.  We begin with a brief review of the most
relevant features of the model.

The HSM is most conveniently formulated by embedding the
one-dimensional chain with periodic boundary conditions into the
complex plane by mapping it onto the unit circle with the spins
located at complex positions
$\eta_\alpha=\exp\!\left(i\frac{2\pi}{N}\alpha\right)$, where $N$
denotes the number of sites and $\alpha=1,\ldots,N$.  The Hamiltonian
is given by
\begin{equation}
  \label{eq:ham}
  H_{\text{HS}}
  =\frac{2\pi^2}{N^2}
  \sum^N_{\alpha<\beta}
  \frac{P_{\alpha\beta}}{\vert \eta_{\alpha}-\eta_{\beta}\vert^2},
\end{equation}
where $P_{\alpha\beta}$ permutes the spins on sites $\alpha$ and
$\beta$.  For the SU(2) model, \eqref{eq:ham} takes the more
familiar form if we substitute $P_{\alpha\beta}
=2\bs{S}_{\alpha}\!\cdot\!\bs{S}_{\beta}+\frac{1}{2}$.
The model is integrable~\cite{talstra-95jpa2369} and possesses a
Yangian symmetry algebra generated by the total spin $\bs{S}$ and the
rapidity operator $\bs{\Lambda}$, which both commute with the
Hamiltonian but do not commute mutually~\cite{haldane-92prl2021}.
 
The ground state ($N=nM$, $M$ integer) for the SU(2) model ($n=2$)
is given by
\begin{equation}
  \label{eq:groundstate}
  \ket{\Psi_0}=P_{\text{G}}\ket{\Psi_{\text{SD}}^N},\quad
  \ket{\Psi_{\text{SD}}^N}\equiv
  \prod_{q\in\mathcal{I}} c_{q\up}^\dagger c_{q\dw}^\dagger\ket{0},
\end{equation}
where the Gutzwiller projector $P_{\text{G}}$ eliminates
configurations with more than one particle on any site 
and the interval $\mathcal{I}$ contains $M$ adjacent momenta.  For
SU($n$), each momenta in $\mathcal{I}$ has to be occupied by $n$
particles with different spins~\cite{kawakami92prbr3191,
  ha-92prb9359}.  As the $N$-particle Slater-determinant state
$\ket{\Psi_{\text{SD}}^N}$ is a spin singlet by construction and
$P_{\text{G}}$ commutes with SU($n$) rotations, $\ket{\Psi_0}$ is an
SU($n$) singlet as well.

A non-orthogonal but complete basis for spin-polarized two-spinon
eigenstates with total momentum $p=-k_1-k_2$ is given by  
\begin{equation}
  \ket{\Psi_{p_1\up,p_2\up}}=P_{\text{G}}\;\!c_{k_1\dw}c_{k_2\dw}
  \!\ket{\Psi_{\text{SD}}^{N+2}}\!,\quad k_1>k_2.
  \label{eq:twospinons}
\end{equation}
These states are not eigenstates, but as $H_{\text{HS}}$ scatters
$k_1$ and $k_2$ in only one direction (increasing $k_1-k_2$), there is
a one-to-one correspondence between these basis states and the exact
eigenstates constructed by superposition.  The total energy of the
eigenstates takes the form \eqref{eq:Lspinonenergy}
if and only if the single-spinon momenta are
shifted with respect to $k_{1,2}$~\cite{greiter-05prb224424,Schuricht-05prb}:
\begin{equation}
  \label{eq:twospinonmomenta}
  p_{1,2}=-k_{1,2}\pm\frac{1}{2n}\frac{2\pi}{N},\quad p_1<p_2.
\end{equation}
The shift can be interpreted as manifestation of the fractional
statistics of the spinons~\cite{haldane91prl937,greiter06}.
For SU(2), an energetically degenerate
two-spinon singlet state $\Lambda^zS^-\ket{\Psi_{p_1\up,p_2\up}}$ with
the same single-spinon momenta exists only for
$p_2-p_1>\frac{1}{2}\frac{2\pi}{N}$, as \eqref{eq:twospinonmomenta} is
annihilated by $\Lambda^zS^-$ for
$p_2-p_1=\frac{1}{2}\frac{2\pi}{N}$~\cite{haldane-92prl2021}.  These
features illustrate that the rules specifying the allowed
single-spinon momenta and spin representations are non-trivial.

\begin{figure}[tb]
\setlength{\unitlength}{8pt}
\begin{picture}(32,8)(1.5,-1)
\linethickness{0.5pt}
\put(3,6){\line(1,0){1}}
\put(3,5){\line(1,0){1}}
\put(3,5){\line(0,1){1}}
\put(4,5){\line(0,1){1}}
\put(3,5){\makebox(1,1){1}}
\put(4.6,5.3){$\otimes$}
\put(6,6){\line(1,0){1}}
\put(6,5){\line(1,0){1}}
\put(6,5){\line(0,1){1}}
\put(7,5){\line(0,1){1}}
\put(6,5){\makebox(1,1){2}}
\put(3,4.4){$\underbrace{\phantom{iiiiiiiiii}}$}
\put(3,2.9){\line(1,0){1}}
\put(3,1.9){\line(1,0){1}}
\put(3,0.9){\line(1,0){1}}
\put(3,0.9){\line(0,1){2}}
\put(4,0.9){\line(0,1){2}}
\put(3,1.9){\makebox(1,1){1}}
\put(3,0.9){\makebox(1,1){2}}
\put(2,-0.5){$S\!=\!0$}
\put(4.6,1.5){$\oplus$}
\put(6,2.4){\line(1,0){2}}
\put(6,1.4){\line(1,0){2}}
\put(6,1.4){\line(0,1){1}}
\put(7,1.4){\line(0,1){1}}
\put(8,1.4){\line(0,1){1}}
\put(6,1.4){\makebox(1,1){1}}
\put(7,1.4){\makebox(1,1){2}}
\put(5.7,-0.2){$S\!=\!1$}
\put(7.6,5.3){$\otimes$}
\put(9,6){\line(1,0){1}}
\put(9,5){\line(1,0){1}}
\put(9,5){\line(0,1){1}}
\put(10,5){\line(0,1){1}}
\put(9,5){\makebox(1,1){3}}
\put(11.6,5.3){$=$}
\put(14,6){\line(1,0){1}}
\put(14,5){\line(1,0){1}}
\put(14,4){\line(1,0){1}}
\put(14,3){\line(1,0){1}}
\put(14,3){\line(0,1){3}}
\put(15,3){\line(0,1){3}}
\put(14,5){\makebox(1,1){1}}
\put(14,4){\makebox(1,1){2}}
\put(14,3){\makebox(1,1){3}}
\put(16.1,5.3){$\oplus$}
\put(18,6){\line(1,0){2}}
\put(18,5){\line(1,0){2}}
\put(18,4){\line(1,0){1}}
\put(18,4){\line(0,1){2}}
\put(19,4){\line(0,1){2}}
\put(20,5){\line(0,1){1}}
\put(18,5){\makebox(1,1){1}}
\put(19,5){\makebox(1,1){2}}
\put(18,4){\makebox(1,1){3}}
\put(17.5,2.2){$S\!=\!\frac{1}{2}$}
\put(21.1,5.3){$\oplus$}
\put(23,6){\line(1,0){2}}
\put(23,5){\line(1,0){2}}
\put(23,4){\line(1,0){1}}
\put(23,4){\line(0,1){2}}
\put(24,4){\line(0,1){2}}
\put(25,5){\line(0,1){1}}
\put(23,5){\makebox(1,1){1}}
\put(23,4){\makebox(1,1){2}}
\put(24,5){\makebox(1,1){3}}
\put(22.5,2.2){$S\!=\!\frac{1}{2}$} 
\put(26.1,5.3){$\oplus$}
\put(28,6){\line(1,0){3}}
\put(28,5){\line(1,0){3}}
\put(28,5){\line(0,1){1}}
\put(29,5){\line(0,1){1}}
\put(30,5){\line(0,1){1}}
\put(31,5){\line(0,1){1}}
\put(28,5){\makebox(1,1){1}}
\put(29,5){\makebox(1,1){2}}
\put(30,5){\makebox(1,1){3}}
\put(28,2.2){$S\!=\!\frac{3}{2}$}
\put(14.5,4.5){\makebox(0,0){\line(1,2){1.8}}}
\put(14.5,4.5){\makebox(0,0){\line(1,-2){1.8}}}
\end{picture}
\caption{Total spin representations of three $S=\frac{1}{2}$ spins
  with Young tableaux. For SU($n$) with $n>2$, the tableaux with three
  boxes on top of each other exists as well.}
\label{fig:youngdiagram}
\end{figure}
We now proceed by stating these rules without further motivating or
even deriving them.  To begin with, the Hilbert space of a system of
$N$ identical SU($n$) spins can be decomposed into representations of
the total spin, which commutes with \eqref{eq:ham} and hence can be
used to classify the eigenstates.  This decomposition can be obtained
using Young tableaux~\cite{Youngtableaux}, as illustrated for three
$S=\frac{1}{2}$ spins in Fig.~\ref{fig:youngdiagram}.  The general
rule is as follows.
For each of the $N$ spins, draw a box numbered consecutively from left
to right.  The representations of SU($n$) are constructed by putting
the boxes together such that the numbers assigned to them increase in
each row from left to right and in each column from top to bottom.
Each tableau indicates symmetrization over all boxes in the same row,
and antisymmetrization over all boxes in the same column.  This
implies that we cannot have more than $n$ boxes on top of each other
for SU($n$) spins.  For SU(2), each tableau corresponds to a spin
$S=\frac{1}{2}(\lambda_1-\lambda_2)$ representation, with $\lambda_i$
the number of boxes in the $i\,\text{th}$ row, and stands for a
multiplet $S^z=-S,\ldots,S$.

\begin{figure}[tb]
\setlength{\unitlength}{8pt}
\begin{picture}(29,6)(0,0)
\linethickness{0.5pt}
\put(5.25,4){\makebox(1,1){$S_{\text{tot}}$}}
\put(21.5,4){\makebox(1,1){$L$}}
\put(24,4.1){$a_1,\dots,a_L$}
\put(0,3){\line(1,0){2}}
\put(0,2){\line(1,0){2}}
\put(0,1){\line(1,0){2}}
\put(0,1){\line(0,1){2}}
\put(1,1){\line(0,1){2}}
\put(2,1){\line(0,1){2}}
\put(0,2){\makebox(1,1){1}}
\put(0,1){\makebox(1,1){2}}
\put(1,2){\makebox(1,1){3}}
\put(1,1){\makebox(1,1){4}}
\put(5.25,1.5){\makebox(1,1){0}}
\put(7,1.8){$\rightarrow$}
\put(9,3){\line(1,0){2}}
\put(9,2){\line(1,0){2}}
\put(9,1){\line(1,0){2}}
\put(9,1){\line(0,1){2}}
\put(10,1){\line(0,1){2}}
\put(11,1){\line(0,1){2}}
\put(9,2){\makebox(1,1){1}}
\put(9,1){\makebox(1,1){2}}
\put(10,2){\makebox(1,1){3}}
\put(10,1){\makebox(1,1){4}}
\put(14,1.8){$\rightarrow$}
\put(16,3){\line(1,0){2}}
\put(16,2){\line(1,0){2}}
\put(16,1){\line(1,0){2}}
\put(16,1){\line(0,1){2}}
\put(17,1){\line(0,1){2}}
\put(18,1){\line(0,1){2}}
\put(16,2){\makebox(1,1){1}}
\put(16,1){\makebox(1,1){2}}
\put(17,2){\makebox(1,1){3}}
\put(17,1){\makebox(1,1){4}}
\put(21.5,1.5){\makebox(1,1){0}}
\put(24,2){\line(1,0){5}}
\multiput(25,1.85)(1,0){4}{\rule{0.5pt}{2pt}}
\end{picture}

\begin{picture}(29,3.5)(0,0)
\linethickness{0.5pt}
\put(0,3){\line(1,0){2}}
\put(0,2){\line(1,0){2}}
\put(0,1){\line(1,0){2}}
\put(0,1){\line(0,1){2}}
\put(1,1){\line(0,1){2}}
\put(2,1){\line(0,1){2}}
\put(0,2){\makebox(1,1){1}}
\put(1,2){\makebox(1,1){2}}
\put(0,1){\makebox(1,1){3}}
\put(1,1){\makebox(1,1){4}}
\put(5.25,1.5){\makebox(1,1){0}}
\put(7,1.8){$\rightarrow$}
\put(9,3){\line(1,0){2}}
\put(9,2){\line(1,0){3}}
\put(10,1){\line(1,0){2}}
\put(9,2){\line(0,1){1}}
\put(10,1){\line(0,1){2}}
\put(11,1){\line(0,1){2}}
\put(12,1){\line(0,1){1}}
\put(9,2){\makebox(1,1){1}}
\put(10,2){\makebox(1,1){2}}
\put(10,1){\makebox(1,1){3}}
\put(11,1){\makebox(1,1){4}}
\put(14,1.8){$\rightarrow$}
\put(16,3){\line(1,0){2}}
\put(16,2){\line(1,0){3}}
\put(17,1){\line(1,0){2}}
\put(16,2){\line(0,1){1}}
\put(17,1){\line(0,1){2}}
\put(18,1){\line(0,1){2}}
\put(19,1){\line(0,1){1}}
\put(16,2){\makebox(1,1){1}}
\put(17,2){\makebox(1,1){2}}
\put(17,1){\makebox(1,1){3}}
\put(18,1){\makebox(1,1){4}}
\put(16.5,1.5){\circle*{0.4}}
\put(18.5,2.5){\circle*{0.4}}
\put(21.5,1.5){\makebox(1,1){2}}
\put(24,2){\line(1,0){5}}
\multiput(25,1.85)(1,0){4}{\rule{0.5pt}{2pt}}
\multiput(25,2)(3,0){2}{\circle*{0.5}}
\put(24.5,0.5){\makebox(1,1){1}}
\put(27.5,0.5){\makebox(1,1){4}}
\end{picture}

\begin{picture}(29,3.5)(0,0)
\linethickness{0.5pt}
\put(0,3){\line(1,0){3}}
\put(0,2){\line(1,0){3}}
\put(0,1){\line(1,0){1}}
\put(0,1){\line(0,1){2}}
\put(1,1){\line(0,1){2}}
\put(2,2){\line(0,1){1}}
\put(3,2){\line(0,1){1}}
\put(0,2){\makebox(1,1){1}}
\put(0,1){\makebox(1,1){2}}
\put(1,2){\makebox(1,1){3}}
\put(2,2){\makebox(1,1){4}}
\put(5.25,1.5){\makebox(1,1){1}}
\put(7,1.8){$\rightarrow$}
\put(9,3){\line(1,0){3}}
\put(9,2){\line(1,0){3}}
\put(9,1){\line(1,0){1}}
\put(9,1){\line(0,1){2}}
\put(10,1){\line(0,1){2}}
\put(11,2){\line(0,1){1}}
\put(12,2){\line(0,1){1}}
\put(9,2){\makebox(1,1){1}}
\put(9,1){\makebox(1,1){2}}
\put(10,2){\makebox(1,1){3}}
\put(11,2){\makebox(1,1){4}}
\put(14,1.8){$\rightarrow$}
\put(16,3){\line(1,0){3}}
\put(16,2){\line(1,0){3}}
\put(16,1){\line(1,0){1}}
\put(16,1){\line(0,1){2}}
\put(17,1){\line(0,1){2}}
\put(18,2){\line(0,1){1}}
\put(19,2){\line(0,1){1}}
\put(16,2){\makebox(1,1){1}}
\put(16,1){\makebox(1,1){2}}
\put(17,2){\makebox(1,1){3}}
\put(18,2){\makebox(1,1){4}}
\put(17.5,1.5){\circle*{0.4}}
\put(18.5,1.5){\circle*{0.4}}
\put(21.5,1.5){\makebox(1,1){2}}
\put(24,2){\line(1,0){5}}
\multiput(25,103.85)(1,0){4}{\rule{0.5pt}{2pt}}
\multiput(27,2)(1,0){2}{\circle*{0.5}}
\put(26.5,0.5){\makebox(1,1){3}}
\put(27.5,0.5){\makebox(1,1){4}}
\end{picture}

\begin{picture}(29,3.5)(0,0)
\linethickness{0.5pt}
\put(0,3){\line(1,0){3}}
\put(0,2){\line(1,0){3}}
\put(0,1){\line(1,0){1}}
\put(0,1){\line(0,1){2}}
\put(1,1){\line(0,1){2}}
\put(2,2){\line(0,1){1}}
\put(3,2){\line(0,1){1}}
\put(0,2){\makebox(1,1){1}}
\put(1,2){\makebox(1,1){2}}
\put(0,1){\makebox(1,1){3}}
\put(2,2){\makebox(1,1){4}}
\put(5.25,1.5){\makebox(1,1){1}}
\put(7,1.8){$\rightarrow$}
\put(9,3){\line(1,0){3}}
\put(9,2){\line(1,0){3}}
\put(10,1){\line(1,0){1}}
\put(9,2){\line(0,1){1}}
\put(10,1){\line(0,1){2}}
\put(11,1){\line(0,1){2}}
\put(12,2){\line(0,1){1}}
\put(9,2){\makebox(1,1){1}}
\put(10,2){\makebox(1,1){2}}
\put(10,1){\makebox(1,1){3}}
\put(11,2){\makebox(1,1){4}}
\put(14,1.8){$\rightarrow$}
\put(16,3){\line(1,0){3}}
\put(16,2){\line(1,0){3}}
\put(17,1){\line(1,0){1}}
\put(16,2){\line(0,1){1}}
\put(17,1){\line(0,1){2}}
\put(18,1){\line(0,1){2}}
\put(19,2){\line(0,1){1}}
\put(16,2){\makebox(1,1){1}}
\put(17,2){\makebox(1,1){2}}
\put(17,1){\makebox(1,1){3}}
\put(18,2){\makebox(1,1){4}}
\put(16.5,1.5){\circle*{0.4}}
\put(18.5,1.5){\circle*{0.4}}
\put(21.5,1.5){\makebox(1,1){2}}
\put(24,2){\line(1,0){5}}
\multiput(25,1.85)(1,0){4}{\rule{0.5pt}{2pt}}
\multiput(25,2)(3,0){2}{\circle*{0.5}}
\put(24.5,0.5){\makebox(1,1){1}}
\put(27.5,0.5){\makebox(1,1){4}}
\end{picture}

\begin{picture}(29,3.5)(0,0)
\linethickness{0.5pt}
\put(0,3){\line(1,0){3}}
\put(0,2){\line(1,0){3}}
\put(0,1){\line(1,0){1}}
\put(0,1){\line(0,1){2}}
\put(1,1){\line(0,1){2}}
\put(2,2){\line(0,1){1}}
\put(3,2){\line(0,1){1}}
\put(0,2){\makebox(1,1){1}}
\put(1,2){\makebox(1,1){2}}
\put(2,2){\makebox(1,1){3}}
\put(0,1){\makebox(1,1){4}}
\put(5.25,1.5){\makebox(1,1){1}}
\put(7,1.8){$\rightarrow$}
\put(9,3){\line(1,0){3}}
\put(9,2){\line(1,0){3}}
\put(11,1){\line(1,0){1}}
\put(9,2){\line(0,1){1}}
\put(10,2){\line(0,1){1}}
\put(11,1){\line(0,1){2}}
\put(12,1){\line(0,1){2}}
\put(9,2){\makebox(1,1){1}}
\put(10,2){\makebox(1,1){2}}
\put(11,2){\makebox(1,1){3}}
\put(11,1){\makebox(1,1){4}}
\put(14,1.8){$\rightarrow$}
\put(16,3){\line(1,0){3}}
\put(16,2){\line(1,0){3}}
\put(18,1){\line(1,0){1}}
\put(17,2){\line(0,1){1}}
\put(16,2){\line(0,1){1}}
\put(18,1){\line(0,1){2}}
\put(19,1){\line(0,1){2}}
\put(16,2){\makebox(1,1){1}}
\put(17,2){\makebox(1,1){2}}
\put(18,2){\makebox(1,1){3}}
\put(18,1){\makebox(1,1){4}}
\put(16.5,1.5){\circle*{0.4}}
\put(17.5,1.5){\circle*{0.4}}
\put(21.5,1.5){\makebox(1,1){2}}
\put(24,2){\line(1,0){5}}
\multiput(25,1.85)(1,0){4}{\rule{0.5pt}{2pt}}
\multiput(25,2)(1,0){2}{\circle*{0.5}}
\put(24.5,0.5){\makebox(1,1){1}}
\put(25.5,0.5){\makebox(1,1){2}}
\end{picture}

\begin{picture}(29,3.5)(0,0)
\linethickness{0.5pt}
\put(0,3){\line(1,0){4}}
\put(0,2){\line(1,0){4}}
\put(0,2){\line(0,1){1}}
\put(1,2){\line(0,1){1}}
\put(2,2){\line(0,1){1}}
\put(3,2){\line(0,1){1}}
\put(4,2){\line(0,1){1}}
\put(0,2){\makebox(1,1){1}}
\put(1,2){\makebox(1,1){2}}
\put(2,2){\makebox(1,1){3}}
\put(3,2){\makebox(1,1){4}}
\put(5.25,1.5){\makebox(1,1){2}}
\put(7,1.8){$\rightarrow$}
\put(9,3){\line(1,0){4}}
\put(9,2){\line(1,0){4}}
\put(9,2){\line(0,1){1}}
\put(10,2){\line(0,1){1}}
\put(11,2){\line(0,1){1}}
\put(12,2){\line(0,1){1}}
\put(13,2){\line(0,1){1}}
\put(9,2){\makebox(1,1){1}}
\put(10,2){\makebox(1,1){2}}
\put(11,2){\makebox(1,1){3}}
\put(12,2){\makebox(1,1){4}}
\put(14,1.8){$\rightarrow$}
\put(16,3){\line(1,0){4}}
\put(16,2){\line(1,0){4}}
\put(16,2){\line(0,1){1}}
\put(17,2){\line(0,1){1}}
\put(18,2){\line(0,1){1}}
\put(19,2){\line(0,1){1}}
\put(20,2){\line(0,1){1}}
\put(16,2){\makebox(1,1){1}}
\put(17,2){\makebox(1,1){2}}
\put(18,2){\makebox(1,1){3}}
\put(19,2){\makebox(1,1){4}}
\put(16.5,1.5){\circle*{0.4}}
\put(17.5,1.5){\circle*{0.4}}
\put(18.5,1.5){\circle*{0.4}}
\put(19.5,1.5){\circle*{0.4}}
\put(21.5,1.5){\makebox(1,1){4}}
\put(24,2){\line(1,0){5}}
\multiput(25,2)(1,0){4}{\circle*{0.5}}
\put(24.5,0.5){\makebox(1,1){1}}
\put(25.5,0.5){\makebox(1,1){2}}
\put(26.5,0.5){\makebox(1,1){3}}
\put(27.5,0.5){\makebox(1,1){4}}
\end{picture}
\caption{Young tableau decomposition and the corresponding spinon
  states for an $S=\frac{1}{2}$ spin chain with $N=4$ sites.  The dots
  represent the spinons.  For SU(2), the spinon momentum numbers $a_i$
  are given by the numbers in the boxes of the same column.  Note that
  $\sum (2S_{\text{tot}}+1)=2^N$.}
  \label{fig:foursitesu2}
\end{figure}
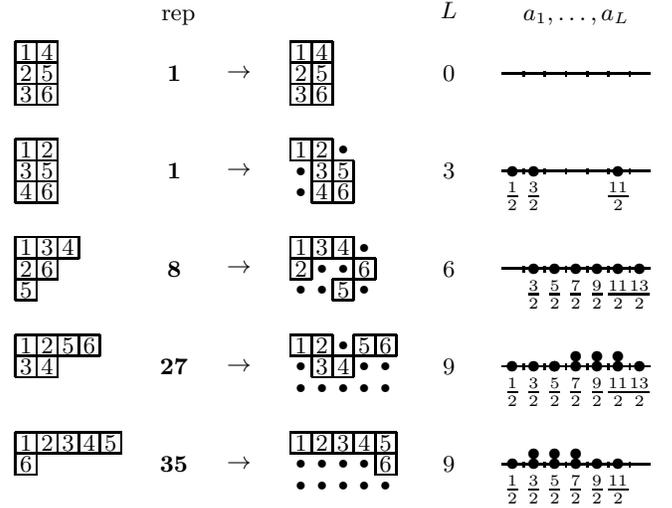
\begin{figure}[tb]
\setlength{\unitlength}{8pt}
\begin{picture}(30,6)(1,0)
\linethickness{0.5pt}
\put(7.9,5){rep}
\put(21,5){\makebox(1,1){$L$}}
\put(25,5){$a_1,\dots,a_L$}

\put(1,1){\line(1,0){2}}
\put(1,2){\line(1,0){2}}
\put(1,3){\line(1,0){2}}
\put(1,4){\line(1,0){2}}
\put(1,1){\line(0,1){3}}
\put(2,1){\line(0,1){3}}
\put(3,1){\line(0,1){3}}
\put(1,3){\makebox(1,1){1}}
\put(1,2){\makebox(1,1){2}}
\put(1,1){\makebox(1,1){3}}
\put(2,3){\makebox(1,1){4}}
\put(2,2){\makebox(1,1){5}}
\put(2,1){\makebox(1,1){6}}
\put(8,2){\makebox(1,1){\bf 1}}
\put(11,2.3){$\rightarrow$}
\put(14,1){\line(1,0){2}}
\put(14,2){\line(1,0){2}}
\put(14,3){\line(1,0){2}}
\put(14,4){\line(1,0){2}}
\put(14,1){\line(0,1){3}}
\put(15,1){\line(0,1){3}}
\put(16,1){\line(0,1){3}}
\put(14,3){\makebox(1,1){1}}
\put(14,2){\makebox(1,1){2}}
\put(14,1){\makebox(1,1){3}}
\put(15,3){\makebox(1,1){4}}
\put(15,2){\makebox(1,1){5}}
\put(15,1){\makebox(1,1){6}}
\put(21,2){\makebox(1,1){0}}
\put(24,2.5){\line(1,0){7}}
\multiput(25,2.35)(1,0){6}{\rule{0.5pt}{2pt}}
\end{picture}

\begin{picture}(30,4.5)(1,0)
\linethickness{0.5pt}
\put(1,4){\line(1,0){2}}
\put(1,3){\line(1,0){2}}
\put(1,2){\line(1,0){2}}
\put(1,1){\line(1,0){2}}
\put(1,1){\line(0,1){3}}
\put(2,1){\line(0,1){3}}
\put(3,1){\line(0,1){3}}
\put(1,3){\makebox(1,1){1}}
\put(1,2){\makebox(1,1){3}}
\put(1,1){\makebox(1,1){4}}
\put(2,3){\makebox(1,1){2}}
\put(2,2){\makebox(1,1){5}}
\put(2,1){\makebox(1,1){6}}
\put(8,2){\makebox(1,1){\bf 1}}
\put(11,2.3){$\rightarrow$}
\put(14,4){\line(1,0){2}}
\put(14,3){\line(1,0){3}}
\put(15,2){\line(1,0){2}}
\put(15,1){\line(1,0){2}}
\put(14,3){\line(0,1){1}}
\put(15,1){\line(0,1){3}}
\put(16,1){\line(0,1){3}}
\put(17,1){\line(0,1){2}}
\put(14,3){\makebox(1,1){1}}
\put(15,3){\makebox(1,1){2}}
\put(15,2){\makebox(1,1){3}}
\put(15,1){\makebox(1,1){4}}
\put(16,2){\makebox(1,1){5}}
\put(16,1){\makebox(1,1){6}}
\put(14.5,2.5){\circle*{0.4}}
\put(14.5,1.5){\circle*{0.4}}
\put(16.5,3.5){\circle*{0.4}}
\put(21,2){\makebox(1,1){3}}
\put(24,2.5){\line(1,0){7}}
\multiput(25,2.35)(1,0){6}{\rule{0.5pt}{2pt}}
\multiput(24.5,2.5)(1,0){2}{\circle*{0.5}}
\put(29.5,2.5){\circle*{0.5}}
\put(24,0.8){\makebox(1,1){$\tfrac{1}{2}$}}
\put(25,0.8){\makebox(1,1){$\tfrac{3}{2}$}}
\put(29,0.8){\makebox(1,1){$\tfrac{11}{2}$}}
\end{picture}

\begin{picture}(30,4.5)(1,0)
\linethickness{0.5pt}
\multiput(1,3)(0,1){2}{\line(1,0){3}}
\put(1,2){\line(1,0){2}}
\put(1,1){\line(1,0){1}}
\multiput(1,1)(1,0){2}{\line(0,1){3}}
\put(3,2){\line(0,1){2}}
\put(4,3){\line(0,1){1}}
\put(1,3){\makebox(1,1){1}}
\put(1,2){\makebox(1,1){2}}
\put(3,3){\makebox(1,1){4}}
\put(2,3){\makebox(1,1){3}}
\put(1,1){\makebox(1,1){5}}
\put(2,2){\makebox(1,1){6}}
\put(8,2){\makebox(1,1){\bf 8}}
\put(11,2.3){$\rightarrow$}
\put(14,4){\line(1,0){3}}
\put(14,3){\line(1,0){4}}
\put(14,2){\line(1,0){1}}
\put(16,2){\line(1,0){2}}
\put(16,3){\line(1,0){1}}
\put(16,1){\line(1,0){1}}
\multiput(14,2)(1,0){2}{\line(0,1){2}}
\multiput(16,1)(0,2){2}{\line(0,1){1}}
\put(17,1){\line(0,1){3}}
\put(18,2){\line(0,1){1}}
\put(14,3){\makebox(1,1){1}}
\put(14,2){\makebox(1,1){2}}
\put(16,3){\makebox(1,1){4}}
\put(15,3){\makebox(1,1){3}}
\put(16,1){\makebox(1,1){5}}
\put(17,2){\makebox(1,1){6}}
\multiput(17.5,1.5)(0,2){2}{\circle*{0.4}}
\multiput(15.5,2.5)(1,0){2}{\circle*{0.4}}
\multiput(14.5,1.5)(1,0){2}{\circle*{0.4}}
\put(21,2){\makebox(1,1){6}}
\put(24,2.5){\line(1,0){7}}
\multiput(25,2.35)(1,0){6}{\rule{0.5pt}{2pt}}
\multiput(25.5,2.5)(1,0){6}{\circle*{0.5}}
\put(25,0.8){\makebox(1,1){$\tfrac{3}{2}$}}
\put(26,0.8){\makebox(1,1){$\tfrac{5}{2}$}}
\put(27,0.8){\makebox(1,1){$\tfrac{7}{2}$}}
\put(28,0.8){\makebox(1,1){$\tfrac{9}{2}$}}
\put(29,0.8){\makebox(1,1){$\tfrac{11}{2}$}}
\put(30,0.8){\makebox(1,1){$\tfrac{13}{2}$}}
\end{picture}

\begin{picture}(30,4.5)(1,0)
\linethickness{0.5pt}
\put(1,4){\line(1,0){4}}
\put(1,3){\line(1,0){4}}
\put(1,2){\line(1,0){2}}
\multiput(1,2)(1,0){3}{\line(0,1){2}}
\multiput(4,3)(1,0){2}{\line(0,1){1}}
\put(1,3){\makebox(1,1){1}}
\put(1,2){\makebox(1,1){3}}
\put(2,2){\makebox(1,1){4}}
\put(2,3){\makebox(1,1){2}}
\put(3,3){\makebox(1,1){5}}
\put(4,3){\makebox(1,1){6}}
\put(8,2){\makebox(1,1){\bf 27}}
\put(11,2.3){$\rightarrow$}
\put(14,4){\line(1,0){2}}
\put(17,4){\line(1,0){2}}
\put(14,3){\line(1,0){5}}
\put(15,2){\line(1,0){2}}
\put(14,3){\line(0,1){1}}
\multiput(15,2)(1,0){3}{\line(0,1){2}}
\multiput(18,3)(1,0){2}{\line(0,1){1}}
\put(14,3){\makebox(1,1){1}}
\put(15,3){\makebox(1,1){2}}
\put(15,2){\makebox(1,1){3}}
\put(16,2){\makebox(1,1){4}}
\put(17,3){\makebox(1,1){5}}
\put(18,3){\makebox(1,1){6}}
\put(14.5,2.5){\circle*{0.4}}
\multiput(14.5,1.5)(1,0){5}{\circle*{0.4}}
\put(16.5,3.5){\circle*{0.4}}
\multiput(17.5,2.5)(1,0){2}{\circle*{0.4}}
\put(21,2){\makebox(1,1){9}}
\put(24,2.5){\line(1,0){7}}
\multiput(25,2.35)(1,0){6}{\rule{0.5pt}{2pt}}
\put(26.5,2.5){\circle*{0.5}}
\multiput(24.5,2.5)(1,0){7}{\circle*{0.5}}
\multiput(27.5,3)(1,0){3}{\circle*{0.5}}
\put(24,0.8){\makebox(1,1){$\tfrac{1}{2}$}}
\put(25,0.8){\makebox(1,1){$\tfrac{3}{2}$}}
\put(26,0.8){\makebox(1,1){$\tfrac{5}{2}$}}
\put(27,0.8){\makebox(1,1){$\tfrac{7}{2}$}}
\put(28,0.8){\makebox(1,1){$\tfrac{9}{2}$}}
\put(29,0.8){\makebox(1,1){$\tfrac{11}{2}$}}
\put(30,0.8){\makebox(1,1){$\tfrac{13}{2}$}}
\end{picture}

\begin{picture}(30,4.5)(1,0)
\linethickness{0.5pt}
\put(1,4){\line(1,0){5}}
\put(1,3){\line(1,0){5}}
\put(1,2){\line(1,0){1}}
\put(1,2){\line(0,1){2}}
\put(2,2){\line(0,1){2}}
\multiput(3,3)(1,0){4}{\line(0,1){1}}
\put(1,3){\makebox(1,1){1}}
\put(2,3){\makebox(1,1){2}}
\put(3,3){\makebox(1,1){3}}
\put(4,3){\makebox(1,1){4}}
\put(5,3){\makebox(1,1){5}}
\put(1,2){\makebox(1,1){6}}
\put(8,2){\makebox(1,1){\bf 35}}
\put(11,2.3){$\rightarrow$}
\put(14,4){\line(1,0){5}}
\put(14,3){\line(1,0){5}}
\put(18,2){\line(1,0){1}}
\put(18,2){\line(0,1){2}}
\put(19,2){\line(0,1){2}}
\multiput(14,3)(1,0){4}{\line(0,1){1}}
\put(14,3){\makebox(1,1){1}}
\put(15,3){\makebox(1,1){2}}
\put(16,3){\makebox(1,1){3}}
\put(17,3){\makebox(1,1){4}}
\put(18,3){\makebox(1,1){5}}
\put(18,2){\makebox(1,1){6}}
\multiput(14.5,2.5)(1,0){4}{\circle*{0.4}}
\multiput(14.5,1.5)(1,0){5}{\circle*{0.4}}
\put(21,2){\makebox(1,1){9}}
\put(24,2.5){\line(1,0){7}}
\multiput(25,2.35)(1,0){6}{\rule{0.5pt}{2pt}}
\multiput(24.5,2.5)(1,0){6}{\circle*{0.5}}
\multiput(25.5,3)(1,0){3}{\circle*{0.5}}
\put(24,0.8){\makebox(1,1){$\tfrac{1}{2}$}}
\put(25,0.8){\makebox(1,1){$\tfrac{3}{2}$}}
\put(26,0.8){\makebox(1,1){$\tfrac{5}{2}$}}
\put(27,0.8){\makebox(1,1){$\tfrac{7}{2}$}}
\put(28,0.8){\makebox(1,1){$\tfrac{9}{2}$}}
\put(29,0.8){\makebox(1,1){$\tfrac{11}{2}$}}
\end{picture}
  \caption{Examples of eigenstates of the SU(3) HSM 
    with $N=6$ sites in terms of colorons.}
  \label{fig:sixsitesu3}
\end{figure}

The main result presented in this Letter is that there is a one-to-one
correspondence between these Young tableaux and the non-interacting
many-spinon states, \ie the eigenstates of the HSM.  The general
principle is illustrated for an SU(2) chain with four sites in
Fig.~\ref{fig:foursitesu2}, and for a few representations of an SU(3)
chain with six sites in Fig.~\ref{fig:sixsitesu3}.  The rule is that
in each Young tableau, we shift boxes to the right such that each box
is below or in the column to the right of the box with the preceding
number.  Each missing box in the resulting, extended tableaux
represents a spinon.  The extended tableaux provide us with the total
spin of each multiplet, which is given by the representation specified
by the original Young tableau, as well as the number $L$ of spinons
present and the individual single-spinon momenta $p_1,\ldots,p_L$.
The latter yield the kinetic energy of the many-spinon state, \ie the
energy of the corresponding eigenstate of the HSM.  We will elaborate
on this now.

To begin with, we introduce a spinon momentum number (SMN) $a_i$ for
each spinon.  For an SU(2) chain, it is simply given by the number in
the box in the same column (see Fig.~\ref{fig:foursitesu2}).  For a
general SU($n$) chain, the SMNs for the spinons in each column are
given by a sequence of numbers (half-integer for $n$ odd, integers for
$n$ even) with integer spacings such that the arithmetic mean equals
the arithmetic mean of the numbers in the boxes of the column.  To
give an example, consider the extended tableau in the second line in
Fig.~\ref{fig:sixsitesu3}.  In the first column, there is only one
box, and the arithmetic mean of the numbers in the boxes is trivially
1.  The SMNs for the two spinons are hence $\frac{1}{2}$ and
$\frac{3}{2}$, as these are integer spaced and have likewise
arithmetic mean 1.  In the last column, there is only one spinon.  The
SMN is given by the arithmetic mean of the two numbers in the boxes,
\ie $(5+6)/2=\frac{11}{2}$.  The individual spinon momenta
corresponding to SMNs $a_i$ are simply given by
\begin{equation}
  \label{eq:singlespinonmom}
  p_i=\frac{2\pi}{N}\:\frac{a_i-\frac{1}{2}}{n},
\end{equation}
which implies $\delta\le p_i\le\frac{2\pi}{n}-\delta$ with 
$\delta=\frac{2\pi}{N}\left(\frac{3}{2n}-\frac{1}{2}\right)\to 0$
for $N\to\infty$.
For SU(2), the total momentum and HS energies of the many-spinon
states are given by \eqref{eq:Lspinonenergy} with
\begin{equation}
  \label{eq:sutwopzeroezero}
  p_0=-\frac{\pi}{2}\:N,\quad 
  E_0=-\frac{\pi^2}{4N},
\end{equation}
and the single-spinon dispersion
\begin{equation}
  \label{eq:sutwoepsilon}
  \epsilon(p)=\frac{1}{2}p\left(\pi-p\right)
    +\frac{\pi^2}{8N^2},
\end{equation}
where we use a convention according to which the ``vacuum'' state
$\ket{\downarrow\downarrow\ldots\downarrow}$ has momentum $p=0$ (and
the empty state $\ket{0}$ has $p=\pi$).  The corresponding formulas
for SU($n$) are
\begin{equation}
  \label{eq:pzeroezero}
  p_0=-\frac{(n\!-\!1)\pi}{n}\:N,\ \
  E_0=-\frac{\pi^2}{12}\!\left(\frac{n\!-\!2}{n}N+\frac{2n\!-\!1}{N}\right)\!,
\end{equation}
and 
\begin{equation}
  \label{eq:epsilon}
  \epsilon(p)=\frac{n}{4}p\left(\frac{2\pi}{n}-p\right)
    +\frac{n^2-1}{12n}\,\frac{\pi^2}{N^2}.
\end{equation}
The simple formalism we just presented provides the complete spectrum
of the HSM.

These results were obtained heuristically, initially
being hardly more than an educated guess at what the equivalent of
Fock space for spinons might be.  It is easy to see that the momentum
spacings for spin polarized spinons predicted by this formalism
reproduce \eqref{eq:twospinonmomenta} for general $n$, and that
spinons of an SU($n$) chain transform under representation
$\bar{\mathbf n}$ of SU($n$)~\cite{bouwknegt-96npb345,Schuricht-05prb}.  It is
also rather easy to see that for SU(2), there is a one-to-one
correspondence between the eigenstates predicted by our formalism and
the known asymptotic Bethe ansatz (ABA)
solution~\cite{kawakami92prb1005,ha-93prb12459} in terms of motifs.
It is further clear that the state counting~\cite{haldane91prl937,
  ha-93prb12459,bouwknegt-96npb345}, \ie the requirement that the
total dimension of the Hilbert space spanned by the many-spinon states
must be $n^N$ for a system consisting of $N$ SU($n$) spins, works out
automatically in the formalism, since the decomposition in 
representations of total spin given by the Young tableaux is complete
and unaffected by our modification of the tableaux.

To establish the correctness of our proposal, we have compared the
spectrum (classified in terms of total spin and momentum quantum
numbers) of the SU(2) and the SU(3) HSM up to 12 sites obtained
numerically by diagonalizing \eqref{eq:ham} with the predictions of
the tableau formalism, and find them identical.  Finally, we have
succeeded recently in showing that the predictions of the formalism
agree with those made by the ABA for general $n$, as we will elaborate
elsewhere~\cite{manuscriptinpreparationGS}.  This is still short of a
rigorous proof as the applicability of the ABA to the model is
heuristic as well, but in light of the success of the ABA solutions
and the numerical work we carried out, we are confident that our
formalism is correct.

\begin{figure}[tb]
\setlength{\unitlength}{8pt}
\begin{picture}(31,7)(-8,6)
\linethickness{0.5pt}
\put(-7.5,10.5){\line(1,0){7}}
\multiput(-6.5,10.35)(1,0){6}{\rule{0.5pt}{2pt}}
\multiput(-6,10.5)(2,0){3}{\circle*{0.5}}
\put(-6.5,8.8){\makebox(1,1){$\tfrac{3}{2}$}}
\put(-4.5,8.8){\makebox(1,1){$\tfrac{7}{2}$}}
\put(-2.5,8.8){\makebox(1,1){$\tfrac{11}{2}$}}
\multiput(19,10)(0,1){2}{\line(1,0){3}}
\put(20,9){\line(1,0){2}}
\put(19,12){\line(1,0){2}}
\put(19,10){\line(0,1){2}}
\multiput(20,9)(1,0){2}{\line(0,1){3}}
\put(22,9){\line(0,1){2}}
\put(19,11){\makebox(1,1){1}}
\put(19,10){\makebox(1,1){2}}
\put(20,11){\makebox(1,1){3}}
\put(20,9){\makebox(1,1){4}}
\put(21,10){\makebox(1,1){5}}
\put(21,9){\makebox(1,1){6}}
\multiput(19.5,9.5)(1,1){3}{\circle*{0.4}}
\put(20,7){\makebox(1,1){\bf 1}}
\put(5,10){\makebox(1,1){$\oplus$}}
\put(5,7){\makebox(1,1){$\oplus$}}
\multiput(7,10)(0,1){3}{\line(1,0){3}}
\put(9,9){\line(1,0){1}}
\multiput(7,10)(1,0){2}{\line(0,1){2}}
\put(9,9){\line(0,1){3}}
\multiput(10,9)(0,2){2}{\line(0,1){1}}
\put(7,11){\makebox(1,1){1}}
\put(7,10){\makebox(1,1){2}}
\put(8,11){\makebox(1,1){3}}
\put(8,10){\makebox(1,1){4}}
\put(9,11){\makebox(1,1){5}}
\put(9,9){\makebox(1,1){6}}
\multiput(7.5,9.5)(1,0){2}{\circle*{0.4}}
\put(9.5,10.5){\circle*{0.4}}
\put(8,7){\makebox(1,1){\bf 8}}
\put(11,10){\makebox(1,1){$\oplus$}}
\put(11,7){\makebox(1,1){$\oplus$}}
\multiput(13,10)(0,1){3}{\line(1,0){3}}
\put(14,9){\line(1,0){1}}
\multiput(13,10)(3,0){2}{\line(0,1){2}}
\multiput(14,9)(1,0){2}{\line(0,1){3}}
\multiput(10,9)(0,2){2}{\line(0,1){1}}
\put(13,11){\makebox(1,1){1}}
\put(13,10){\makebox(1,1){2}}
\put(14,11){\makebox(1,1){3}}
\put(14,9){\makebox(1,1){4}}
\put(15,11){\makebox(1,1){5}}
\put(15,10){\makebox(1,1){6}}
\multiput(13.5,9.5)(2,0){2}{\circle*{0.4}}
\put(14.5,10.5){\circle*{0.4}}
\put(14,7){\makebox(1,1){\bf 8}}
\put(17,10){\makebox(1,1){$\oplus$}}
\put(17,7){\makebox(1,1){$\oplus$}}
\multiput(1,10)(0,1){3}{\line(1,0){3}}
\multiput(1,10)(1,0){4}{\line(0,1){2}}
\put(1,11){\makebox(1,1){1}}
\put(1,10){\makebox(1,1){2}}
\put(2,11){\makebox(1,1){3}}
\put(2,10){\makebox(1,1){4}}
\put(3,11){\makebox(1,1){5}}
\put(3,10){\makebox(1,1){6}}
\multiput(1.5,9.5)(1,0){3}{\circle*{0.4}}
\put(2,7){\makebox(1,1){$\bar{\bf 10}$}}
\end{picture}

\begin{picture}(31,6)(-8,1)
\linethickness{0.5pt}
\put(-7.5,3.5){\line(1,0){7}}
\multiput(-6.5,3.35)(1,0){6}{\rule{0.5pt}{2pt}}
\multiput(-7,3.5)(1,0){2}{\circle*{0.5}}
\put(-2,3.5){\circle*{0.5}}
\put(-7.5,1.8){\makebox(1,1){$\tfrac{1}{2}$}}
\put(-6.5,1.8){\makebox(1,1){$\tfrac{3}{2}$}}
\put(-2.5,1.8){\makebox(1,1){$\tfrac{11}{2}$}}
\put(7,5){\line(1,0){2}}
\put(7,4){\line(1,0){3}}
\multiput(8,2)(0,1){2}{\line(1,0){2}}
\put(7,4){\line(0,1){1}}
\multiput(8,2)(1,0){2}{\line(0,1){3}}
\put(10,2){\line(0,1){2}}
\put(7,4){\makebox(1,1){1}}
\put(8,4){\makebox(1,1){2}}
\put(8,3){\makebox(1,1){3}}
\put(8,2){\makebox(1,1){4}}
\put(9,3){\makebox(1,1){5}}
\put(9,2){\makebox(1,1){6}}
\put(9.5,4.5){\circle*{0.4}}
\multiput(7.5,2.5)(0,1){2}{\circle*{0.4}}
\put(8,0){\makebox(1,1){\bf 1}}
\put(5,3){\makebox(1,1){$\oplus$}}
\put(5,0){\makebox(1,1){$\oplus$}}
\multiput(2,2)(0,1){2}{\line(1,0){1}}
\put(3,3){\line(1,0){1}}
\multiput(1,4)(0,1){2}{\line(1,0){3}}
\put(1,4){\line(0,1){1}}
\multiput(2,2)(1,0){2}{\line(0,1){3}}
\put(4,3){\line(0,1){2}}
\put(1,4){\makebox(1,1){1}}
\put(2,4){\makebox(1,1){2}}
\put(2,3){\makebox(1,1){3}}
\put(2,2){\makebox(1,1){4}}
\put(3,4){\makebox(1,1){5}}
\put(3,3){\makebox(1,1){6}}
\put(1.5,3.5){\circle*{0.4}}
\multiput(1.5,2.5)(2,0){2}{\circle*{0.4}}
\put(2,0){\makebox(1,1){\bf 8}}
\put(12.5,3.5){\line(1,0){5}}
\multiput(13.5,3.35)(1,0){4}{\rule{0.5pt}{2pt}}
\multiput(14,3.5)(1,0){3}{\circle*{0.5}}
\put(13.5,1.8){\makebox(1,1){$\tfrac{7}{2}$}}
\put(14.5,1.8){\makebox(1,1){$\tfrac{9}{2}$}}
\put(15.5,1.8){\makebox(1,1){$\tfrac{11}{2}$}}
\multiput(19,2)(0,1){2}{\line(1,0){1}}
\put(21,3){\line(1,0){1}}
\multiput(19,4)(0,1){2}{\line(1,0){3}}
\multiput(19,2)(1,0){2}{\line(0,1){3}}
\multiput(21,3)(1,0){2}{\line(0,1){2}}
\put(19,4){\makebox(1,1){1}}
\put(19,3){\makebox(1,1){2}}
\put(19,2){\makebox(1,1){3}}
\put(20,4){\makebox(1,1){4}}
\put(21,4){\makebox(1,1){5}}
\put(21,3){\makebox(1,1){6}}
\put(20.5,3.5){\circle*{0.4}}
\multiput(20.5,2.5)(1,0){2}{\circle*{0.4}}
\put(20,0){\makebox(1,1){\bf 8}}
\end{picture}
\caption{Examples of allowed SU(3) spin representations for 
given SMNs 
in a chain with $N=6$ sites.}
\label{fig:threecolorons}
\end{figure}
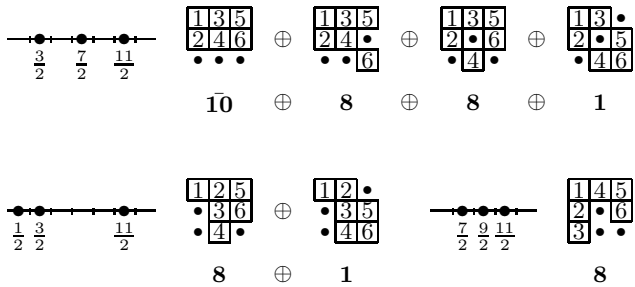
Remarks: (a) For the polarized two-spinon states, there is a
correspondence between the fermion occupations in the basis states
\eqref{eq:twospinons} before projection and the tableau
representations (see \eg the $S_{\text{tot}}=1$ tableaux in
Fig.~\ref{fig:foursitesu2}).  We conjecture that such a connection
exists in general.  If so, it would be desirable to have general rules
for the construction of the required total spin representations with
the fermions annihilated before projection such that the basis states
with the same single-spinon momenta corresponding to different
tableaux are orthogonal.

(b) The tableau formalism implies that the momenta of the spinons, as
compared to the momenta of the fermions before projection, are shifted
by $\frac{2\pi}{N}\frac{1}{2n}$ 
towards each other if they are in different columns of the tableau,
and by the same amount away from each other if they are in the same
column.  (Since the ``bare'' momenta of the fermions are identical in
the latter case, it is not possible to shift them towards each other.)
The statistics of the spinons is hence that of fermions shifted by a
statistical angle $\Delta\theta=\pi/n$~\cite{greiter06}.  This
implies
$\theta=\pi\left(1-\frac{1}{n}\right)$~\cite{
  bouwknegt-99npb501, Schuricht-05prb} for spin-polarized spinons.

(c) The formalism can be used to calculate thermodynamic quantities.
In this regard, it provides an alternative to the ``freezing trick''
method of Sutherland and Shastry~\cite{sutherland-93prl5}.

(d) The formalism provides, as a byproduct, the general rules which
representations are possible for a given set of spinons with given 
single-particle momenta (see Fig.~\ref{fig:threecolorons} for examples).
These rules may also be interpreted in the framework of Yangian
representation theory~\cite{chari-90lem267}. 

(e) Since the low energy physics of the HSM is described by the
$\text{SU}(n)$ level $k=1$ Wess-Zumino-Witten
model~\cite{DiFrancescoMathieuSenechal97}, the rules for combining
representations of spinons one may deduce form our formalism will
apply to the quasiparticles of this theory as well.  This connection
has been exploited by Bouwknegt and
Schoutens~\cite{bouwknegt-96npb345,bouwknegt-99npb501}, who obtained a
significant body of results for SU($n$) spin chains from conformal
field theories.

In conclusion, we have established a one-to-one correspondence between
the Young tableaux classifying the total spin representations of $N$ 
spins and the exact eigenstates of the the Haldane-Shastry model for a
chain with $N$ sites.  This correspondence allows us to label the
many-spinon eigenstates in terms of their single-spinon momenta, which
are spaced according to highly non-trivial rules.  Since the spinons
in the HSM are free in the sense that they only interact through their
fractional statistics, the tableau formalism introduced here provides
a general construction principle of the free spinon Hilbert space, the
analog of a Fock space representation for spinons.

One of us (DS) was supported by the German Research Foundation (DFG)
through GK 284.


\end{document}